\documentclass[preprint,12pt]{elsarticle}

\usepackage{amssymb}

\usepackage{graphicx}
\usepackage{subcaption}
\usepackage{booktabs}
\usepackage{array}
\journal{Ad Hoc Networks}
\begin{document}

\begin{frontmatter}



\title{Eventually-Consistent Federated Scheduling for Data Center Workloads}


\author[inst1]{Meghana Thiyyakat}

\affiliation[inst1]{organization={Department of Computer Science and Engineering\\ PES University},
            city={Bangalore},
            country={India}}

\author[inst1]{Subramaniam Kalambur}
\author[inst1]{Rishit Chaudhary}
\author[inst1]{Saurav G Nayak}
\author[inst1]{Adarsh Shetty}
\author[inst2]{Dinkar Sitaram}

\affiliation[inst2]{organization={Cloud Computing Innovation Council of India},
            city={Bangalore},
            country={India}}

\begin{abstract}
Data center schedulers operate at unprecedented scales today to accommodate the growing demand for computing and storage power. The challenge that schedulers face is meeting the requirements of scheduling speeds despite the scale. To do so, most scheduler architectures use parallelism. However, these architectures consist of multiple parallel scheduling entities that can only utilize partial knowledge of the data center's state, as maintaining consistent global knowledge or state would involve considerable communication overhead. The disadvantage of scheduling without global knowledge is sub-optimal placements—tasks may be made to wait in queues even though there are resources available in zones outside the scope of the scheduling entity's state. This leads to unnecessary queuing overheads and lower resource utilization of the data center.

In this paper, extend our previous work on Megha, a federated decentralized data center scheduling architecture that uses eventual consistency. The architecture utilizes both parallelism and an eventually-consistent global state in each of its scheduling entities to make fast decisions in a scalable manner. In our work, we compare Megha with 3 scheduling architectures -- Sparrow, Eagle, and Pigeon, using simulation. We also evaluate Megha's prototype on a 123-node cluster and compare its performance with Pigeon's prototype using cluster traces. The results of our experiments show that Megha consistently reduces delays in job completion time when compared to other architectures.
\end{abstract}


\end{frontmatter}


\section{Introduction}
\label{sec:introduction}
The rise in the usage of deep learning methods and the increasing size of datasets have created a need for extremely large amounts of computing power. Older task scheduling frameworks, such as Hadoop \cite{shvachko2010hadoop, sandholm2010dynamic}, have been unable to scale to the required number of worker nodes \cite{vavilapalli2013apache, curino2019hydra}. Consequently, several new frameworks that focus on scalability have been developed. In most frameworks, scalability has been implemented by introducing parallelism into the architecture \cite{ousterhout2013sparrow, delgado2016job, wang2019pigeon}, thereby increasing scheduling throughput. These architectures use multiple scheduling entities, each of which has access to up-to-date information on resource availability in some parts of the data center (DC). The drawback of this approach is that the scheduling entities lack access to global resource availability knowledge or the global state of the DC, resulting in tasks being queued even when there are available workers elsewhere in the DC.

In this paper, we present an extended version of our previous work on Megha \cite{thiyyakat2023megha}, a federated, eventually-consistency-based scheduler that uses a decentralized global state to make near-optimal scheduling decisions. We use delay in job completion time (JCT) as the metric for comparing Megha with Sparrow \cite{ousterhout2013sparrow}, Eagle \cite{delgado2016job}, and Pigeon \cite{wang2019pigeon}. We also study Megha's performance under different loads and DC sizes. In addition to simulation runs, we also evaluate Megha by deploying its prototype on 3 Kubernetes clusters. Our simulation and prototype results demonstrate that Megha's architecture and algorithm improve the overall JCT, at scale.

\section{Background}
\label{sec:background}
A data center (DC) is a physical building that houses computing, storage, and networking resources along with supporting components such as security and electrical systems. The DC's scheduler acts as its operating system (OS). It monitors the health of its resources \cite{vavilapalli2013apache}, assigns resources to tasks, handles failures in the DC (both resource \cite{ousterhout2013sparrow, wang2019pigeon} and task), and enforces fairness \cite{schwarzkopf2013omega, boutin2014apollo, burns2016borg, curino2019hydra}. Like an OS, the DC also needs to present a simple, unified interface to the resource pool \cite{curino2019hydra} and arbitrate access to these resources among different users \cite{vavilapalli2013apache}. The primary responsibility of the scheduler, known as task allocation \cite{jiang2015survey}, is to map tasks to resources. Resources are allocated to tasks as \textit{resource units} \cite{ousterhout2013sparrow, wang2019pigeon}. Resource units could simply be a logical bundle of resources or resources isolated and protected from the rest of the system using technologies such as virtualization \cite{protean} or containerization \cite{burns2016borg}. The scheduler maps these units to tasks while optimizing certain objectives.

\subsection{DC Workloads}

The workload that the DC scheduler receives is heterogeneous \cite{reiss2012heterogeneity, delgado2015hawk, liu2022fregata}. The workloads consist of jobs ranging from short, interactive jobs to long-running batch jobs. The DC's resource pool is shared by the jobs in the workload. The motivation to collocate multiple jobs on the same infrastructure, instead of using dedicated infrastructure for each job, is higher resource utilization. Higher resource utilization in the DC leads to higher ROI. Google reports that having different resource pools for production and non-production workloads would increase the number of machines required by 20\%-30\% \cite{burns2016borg}. Jobs of different types have different characteristics and requirements. Typically, short jobs are latency-sensitive and have higher priority. Long jobs are resource-intensive (requiring optimal load balancing), but latency-tolerant \cite{liu2022fregata}.

Cluster traces published by big data companies record normalized workload and infrastructure parameters while obfuscating sensitive information. These traces have been extensively analyzed \cite{reiss2012heterogeneity, clusterdata:Elsayed2017, clusterdata:Sebastio2018, guo2019limits, clusterdata:Tirmazi2020} and have been used to generate workloads to measure and compare the performance of frameworks \cite{ousterhout2013sparrow, delgado2015hawk, delgado2016job, hao2017pcssampler, wang2019pigeon}. Google has published two cluster traces, one in 2011 \cite{clusterdata:Wilkes2011} and the second in 2019 \cite{clusterdata:Wilkes2020}. Jobs are submitted to Google's scheduler, Borg \cite{burns2016borg}, with details such as the resources required, the binaries to be deployed, and the priority of the job. Jobs also specify an upper limit for resources to help the scheduler determine whether a task instance can be accommodated on a machine. An analysis of the two traces \cite{clusterdata:Tirmazi2020} shows that the job submission rate has increased by 3.7x since 2011, and the number of tasks needing to be scheduled by 7x. This demonstrates that the load on DC schedulers has increased considerably. Hydra \cite{curino2019hydra}, Microsoft's DC scheduler, handles a workload consisting of over 500,000 jobs (billions of tasks) daily, from over 5 frameworks, requiring the scheduler to support a rate of over 40,000 scheduling decisions per second.
\subsection{DC Scheduler Architectures} \label{sec:architectures}
When cluster sizes were smaller, schedulers had monolithic architectures. However, due to the lack of parallelism in the architecture, they were unable to scale to the increasing data center sizes. Consequently, two-tiered architectures \cite{hindman2011mesos,vavilapalli2013apache} were developed, which separated resource allocation from functions specific to applications such as task scheduling and failure handling. These architectures, however, continued to use a single centralized job scheduler, which became a bottleneck as the workloads became even more demanding.

\subsubsection{Centralized Architectures}

The earliest schedulers that tried to solve the scalability problem were YARN \cite{vavilapalli2013apache} and Mesos \cite{hindman2011mesos}, which had a Centralized Architecture. In a centralized architecture, the scheduler has a single scheduling entity that is responsible for allocating resources to tasks based on some criteria. The resource allocator has access to the up-to-date status of all the resources in the DC. The scheduler is divided into a single coordinator and multiple agent components. The agents run on all worker nodes and periodically report to the coordinator on the health and status of the resources in the worker nodes.

\subsubsection{Distributed Architectures}
Due to the bottleneck arising from using a single scheduling entity, centralized architectures were replaced by schedulers with Distributed Architectures. These schedulers had multiple scheduling entities working in parallel and could achieve greater scheduling speed and throughput. The only pitfall in this architecture was that none of the scheduling entities had complete knowledge of the resource availability in the DC. This resulted in poor task placement decisions under high loads.

Our scheduler, Megha, has been compared with a popular distributed scheduler, Sparrow \cite{ousterhout2013sparrow}, which was developed at Berkeley with the primary goal of scalability with high scheduling throughput and low scheduling latency. The architecture consists of multiple autonomous schedulers or resource allocators, and worker machines that execute the tasks. Each worker machine has an associated task queue into which the schedulers insert tasks. When a job arrives, the scheduler uses batch sampling and late binding to map the tasks to nodes. Instead of probing \texttt{d} nodes per task, the scheduler probes \texttt{d x n} nodes per job, where \texttt{n} is the number of tasks in the job. Instead of using the queue length as a heuristic to measure waiting time at a node, the probe adds a reservation for one task at the end of the worker's queue. When the reservation reaches the front of the queue, the worker sends an RPC to the scheduler. The scheduler dispatches the \texttt{n} tasks to the first \texttt{n} worker nodes to respond.

\subsubsection{Hybrid Architectures}
DC workloads are heterogeneous, consisting of a mixture of short and long jobs. Scheduling both types of jobs using a centralized scheduler would lead to unacceptable delays in the short jobs. On the other hand, scheduling long jobs using distributed schedulers would result in sub-optimal placements and resource utilization. Therefore, hybrid schedulers combine both types of schedulers to cater to the needs of both short and long jobs.

While centralized schedulers make decisions using the entire global state of the DC and therefore make more optimal scheduling decisions, in distributed architectures, multiple autonomous schedulers operate in parallel to increase throughput, but with only a partial state of the system. To combine the benefits of both centralized and distributed schedulers, hybrid architectures employ both types of schedulers. Based on the performance requirements of the job, the job's tasks are directed either to one of the distributed schedulers or the single centralized scheduler. However, due to a lack of coordination between the different entities, global fairness policies could not be enforced in these architectures. Moreover, under high loads, task placement continued to be sub-optimal \cite{wang2019pigeon}.

In this paper, Megha has been compared with \textbf{Eagle} \cite{delgado2016job}, a hybrid scheduler. Eagle assumes that in a DC workload, the number of long jobs is few, but they consume a large portion of the resources, whereas there are a large number of short jobs that are latency-sensitive but require a smaller fraction of the DC's resources. Eagle also uses an estimate of the task execution times to classify the jobs as short or long based on a given threshold. The basis for this assumption is that many jobs are recurring and execute on similar input, making it easier to estimate job duration from previous runs. Eagle divides the DC into a short partition and a long partition.

One of the major sources of delay in short jobs is head-of-line blocking, which occurs when a short task gets stuck behind a long task in a worker's queue. Eagle proactively eliminates this issue with the help of Succinct State Sharing (SSS). The SS is a bit vector that is sent to worker nodes by the centralized scheduler, indicating the nodes on which long jobs have been scheduled. Worker nodes that are currently running long tasks respond to probes sent by distributed schedulers with their SS. The probe is considered a rejected probe. The distributed scheduler picks the SS with the most recent timestamp to update its view of the DC and re-sends the rejected probes to only those worker nodes that do not have a long job running on them according to the SS. Due to the staleness of the most recent SS sent to the distributed scheduler, there is a chance that probes may get rejected once more. In this case, the probes are sent to workers in the short partition picked at random. Eagle also uses Sticky Batch Probing to ensure faster job completion. Once a task completes its execution on a worker, instead of relinquishing the worker to a random task, the worker checks to see if there are tasks remaining in the same job and executes them instead. This reduces the number of jobs running on the DC at any instant and thereby improves job completion time (Little's Law \cite{little1961proof}).

\subsubsection{Federated Architectures}
The drawbacks in earlier architectures led to the adoption of schedulers with Federated Architectures \cite{curino2019hydra,wang2019pigeon}. These schedulers use a decentralized approach to distribute the scheduling responsibility across multiple tiers, facilitating superior scalability and high scheduling speeds. Federated schedulers divide the DC into smaller clusters. Each cluster is managed by a local scheduler that can function nearly autonomously. The local scheduler takes care of responsibilities such as task placement, monitoring the state of the cluster, ensuring high availability, etc. Global policy is decided by the top-level components and pushed out to the local schedulers for enforcement. Due to the divide-and-conquer approach employed, federated schedulers are scalable and fast.

\textbf{Pigeon} \cite{wang2019pigeon} is a federated scheduler whose architecture closely resembles Megha's. It organizes the DC's nodes into smaller groups, each governed by a group coordinator. At the top level are distributed schedulers that assign incoming tasks to different coordinators. This load balancing is performed without global knowledge and does not take into account the type of job that the tasks belong to. Pigeon also prioritizes short jobs over long jobs with two weighted fair queues and by limiting the number of nodes on which tasks from long jobs can run. When a job arrives at a distributor, its tasks are distributed evenly to all the coordinators. Each coordinator either assigns the task to an available worker node or inserts it into a low or high-priority queue based on whether the job is long or short, respectively. In each group, certain worker nodes are reserved to execute only high-priority tasks. Therefore, while high-priority tasks can run on all the worker nodes in the group, low-priority tasks cannot run on worker nodes reserved for high-priority tasks. When the coordinator receives a low-priority task, it checks its list of available low-priority workers. If a worker node is available, the task is assigned to the worker. If not, the task is inserted into the low-priority queue. On the other hand, when the coordinator receives a high-priority task, it first checks the list of low-priority workers available and assigns the task to an available worker. If such a worker is unavailable, the coordinator checks for nodes available in the worker nodes reserved for high-priority tasks. If a worker node is found, the task is assigned to it; otherwise, the task is inserted into the high-priority queue. The coordinator picks a task to schedule from the queues based on weighted fair queuing. If $W$ is the weight, then for every $W$ high-priority task, the scheduler must pick one low-priority task. This prevents starvation of low-priority tasks while giving preference to high-priority tasks.

These different architectures have evolved over time to address the scalability, efficiency, and fairness challenges posed by large data center workloads. Each architecture has its strengths and weaknesses, and the choice of architecture depends on the specific requirements and constraints of the data center environment.

\subsection{DC Scheduler Metrics}
 This section describes the various metrics used to assess and compare DC scheduler performance.

\subsubsection{Delays}

One of the main objectives of a scheduler is to facilitate optimal completion times for jobs, that is, to schedule jobs with minimal delay. A job is a collection of tasks and is said to have been completed only when all its tasks have been completed. We define job completion time as follows:
\begin{equation}
    JCT_{i} = JRT_{i} - JST_{i}
\end{equation}
where $JCT_i$ is the time taken for all the tasks in the job to complete execution,and $JRT_i$ is the wall clock time at which the job responds saying it has completed execution, and $JST_{i}$ is the job submission time or the time at which the job was submitted to the scheduler.
If $T = {T_{i,1},T_{i,2}...,T_{i,N}}$ is the set of all tasks of a job $i$, then $JCT_{i} \geq T_{i,max}$, where $T_{i,max} \in T$ is the task which executes for the longest time. The delay $d^{job}_i$ in the completion time of a job $i$, $JCT_i$, is:
\begin{equation}\label{eq:job_delay}
    d^{job}_i = JCT_i - IdealJCT_i
\end{equation}
$IdealJCT_i$ is the job completion time of $i$ when scheduled by an omniscient scheduler on a DC of infinite size. Task completion time, $TCT_{i,j}$ of a task $j$, belonging to job $i$, is the time between $JST_i$ and $TRT_{i,j}$ which is the time at which the task responds saying it has completed execution:
\begin{equation}
    TCT_{i,j} = TRT_{i,j} - JST_{i}
\end{equation}
The delay in task completion
time of task $j$, $d^{task}_{i,j}$ is:
\begin{equation}
    d^{task}_{i,j} = TCT_{i,j} - IdealTET_{i,j}
\end{equation}
$IdealTET_{i,j}$ is the ideal task execution time or the amount of time the task would execute on a worker node under ideal conditions, that is, in the absence of throttling, interference, etc. The delay in task completion time,$d^{task}_{i,j}$, can be further split into the following components:
\begin{equation}
     d^{task}_{i,j} = d\_queue^{scheduler}_{i,j} + d\_proc
     _{i,j} + d\_comm
     _{i,j} + d\_queue^{worker}_{i,j} + d\_exec_{i,j} 
\end{equation}

$d\_queue^{scheduler}_{i,j}$ is the amount of time $T_{i,j}$ spends waiting in a scheduler's job queue. $d\_proc_{i,j}$ is the amount of time the scheduler takes to process the task's request by searching for resources using a global or partial state knowledge. $d\_comm_{i,j}$ is the time taken by the components in the scheduler to exchange the messages necessary to execute a task. $d\_queue^{worker}_{i,j}$  is the amount of time the task spends queued at a worker node. $d\_exec_{i,j}$ is the execution delay experienced by a task due to less-than-ideal execution conditions. One or more delay components may be missing. For instance, for a centralized scheduler that does not employ worker-side queuing, the $d\_queue^{worker}_{i,j}$ component would be irrelevant. Similarly, Sparrow \cite{ousterhout2013sparrow} does not have scheduler-side queues, hence the $d\_queue^{scheduler}_{i,j}$ is not applicable. Some components can be an aggregate of multiple smaller delays. The $d\_comm_{i,j}$ component in Pigeon \cite{wang2019pigeon} can be split into the communication delays incurred when sending the task from the scheduler to the coordinator, and then sending the task from the coordinator to the worker. 

In some functions, the delays overlap, and cannot be blindly aggregated. This is true in most schedulers that use random-sample-based probing \cite{ousterhout2013sparrow,delgado2015hawk,karanasos2015mercury,delgado2016job,hao2017pcssampler,thinakaran2017phoenix} where reservations are queued at more than one worker node. Since the reservations are made in parallel, the queuing delays of the reservations overlap. Therefore, the $d\_queue^{worker}_{i,j}$ in this case, would be equal to just the queuing delay at the worker node that is the first to respond to the scheduler, indicating that it has resources to run the task. 

Some schedulers try to reduce the execution delay component, $d\_exec_{i,j}$, by making interference-aware task placements. Frameworks such as \cite{delimitrou2013paragon}, Quasar \cite{delimitrou2014quasar} and Bubble-flux\cite{yang2013bubble} use profiling to understand the interference caused by each task, and also its sensitivity to interference from other co-located tasks. Others \cite{lo2015heracles,perfiso,li2019pine} improve the resource isolation of tasks at the OS level using kernel features.

\subsubsection{Scalability}

Recent work \cite{curino2019hydra} has revealed that schedulers of today are required to make over 40k scheduling decisions per second (SDPS). Additionally, they must be able to manage clusters with tens of thousands of worker nodes \cite{boutin2014apollo,burns2016borg, curino2019hydra}. Data analytic workloads consisting of predominantly short jobs (tasks of 100ms duration) running on a cluster of ten thousand machines may require speeds over 1 million SDPS \cite{ousterhout2013sparrow}. Older architectures such as YARN (scalable to up to 4k worker nodes) were unable to keep up with scheduling speeds and the cluster sizes required to support today's workloads \cite{curino2019hydra} leading to the development of federated and hierarchical architectures. Hydra is reported to be able to achieve 30k-40k SDPS and scale to 50k worker nodes. Pigeon's architecture \cite{wang2019pigeon} was evaluated using simulation for cluster sizes of up to 18k worker nodes, and the results show that it improves upon hybrid schedulers such as Eagle \cite{delgado2016job} as well as centralized schedulers.

 \subsubsection{Resource Utilization}

 Idle resources result in poor ROI. Therefore, the scheduler must balance resource utilization with scheduling overheads and performance. Finding worker nodes with sufficient resources for a task while optimizing the number of worker nodes on which the workload is consolidated is essentially a bin-packing problem, which is NP-hard. Therefore, it is computationally very expensive and can lead to large scheduling overheads. Schedulers instead use heuristics to find a near-optimal task placement. Most schedulers use a greedy solution of finding any available worker for a task, or if that fails, placing a task in a worker's queue with the least estimated wait time. The accuracy of the heuristic depends on the quality and quantity of the resource status information available to the scheduler. When a scheduler employs partial state knowledge, such as the distributed scheduling entities in Sparrow, scheduling decisions may be sub-optimal. This is because, under high load, the probability of probes finding available worker nodes is smaller, resulting in unnecessary queuing at busy worker nodes. Eagle uses SSS to aid in task placement by giving the distributed schedulers a (possibly stale) hint about the locations of long tasks. Sub-optimal load balancing during task placements can also lead to idle resources and unnecessary queuing. To avoid that, using the law of large numbers, Pigeon \cite{wang2019pigeon} distributes all the tasks in a job evenly to all the coordinators rather than to individual workers for even load balancing across groups. However, once a task is sent to a group's coordinator, it is restricted to the resources in the group. Therefore, if there are idle resources in other groups, and a group is overwhelmed due to stragglers, tasks from one coordinator's queues cannot be migrated to another's.

  When a task is queued unnecessarily at a scheduler or worker even though there are available resources in the DC, the delay in jobs increases, and resource utilization decreases due to idle resources. Reducing unnecessary queuing delay in jobs also results in better resource utilization of the DC. Megha makes use of decentralized global state in GMs to find available worker nodes easily and avoid unnecessary queuing.

\section{Megha}

Megha is a federated scheduler that relies on eventually-consistent global state knowledge of the entire DC to make fast and near-optimal task placement decisions. 

\label{sec:megha}
\subsection{Architecture}

\begin{figure}[]
    \centering
       \includegraphics[width=\linewidth]{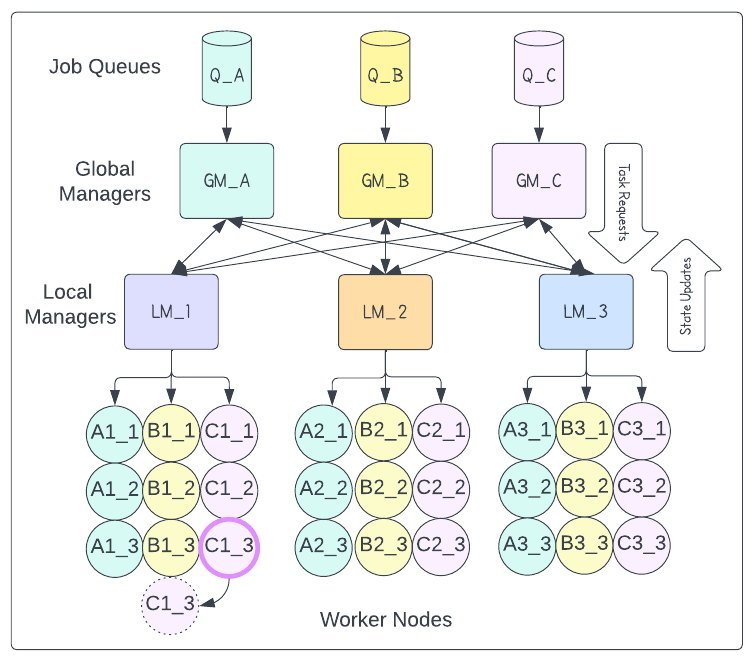}
    \caption{Megha's Architecture}
    \label{fig:Megha}
\end{figure}

Megha's architecture consists of two types of scheduling entities- Global Managers (GM) and Local Managers (LM). Like Pigeon, Megha divides the DC into smaller clusters. Each cluster is monitored and managed autonomously by an LM. Jobs are distributed evenly across GMs and are inserted into the GMs' Job Queues. The GMs gather state information from each LM about their respective clusters and use it collectively to make task placement decisions. Fig. ~\ref{fig:Megha} shows Megha deployed in a 3GM-3LM and 27 worker nodes configuration. The GMs have been labeled as $GM\_i$, where $i \in \{A,B,C\}$, the Job Queue for a GM, $GM\_i$, has been labeled $Q\_i$, the LMs have been labeled as $LM\_j$, where $j \in \{1,2,3\}$. The clusters under each LM have been further divided into \textit{partitions}. Each GM is assigned one partition under each cluster. In Fig.~\ref{fig:Megha}, worker nodes have been labeled as $ij\_n$, where $n \in {1,2,3}$. The values for $ij$ are the same for all worker nodes in a partition. For instance, worker node $A1\_1$, $A1\_2$, and $A1\_3$, with $i=A$ and $j=1$, are worker nodes of a partition under $LM\_1$ belonging to $GM\_A$. The partitions belonging to a GM are known as the GM's \textit{internal partitions}, other partitions are referred to as the GM's \textit{external partitions}. 

Megha combines the benefits of both centralized and distributed architectures - the GMs use global state knowledge of the DC while making task placement decisions and, multiple GMs operate in parallel. However, because of the prohibitive coordination overheads involved in maintaining a consistent decentralized global state in all the GMs, Megha relaxes the consistency requirement. Instead, GMs use request validation and aperiodic LM state updates along with the eventually-consistent, decentralized global state to make near-optimal task placement decisions. 

\subsection{Task Scheduling}
Jobs are evenly distributed across GMs. When a job arrives at a GM, it is inserted into the GM's job queue. When resources are available, the GM reads a job from the front of its job queue and maps each task in the job to an available worker node. Each GM stores, locally, the global state knowledge of the DC aggregated from periodic and aperiodic status updates from the LMs. This global state is used by the GMs to find available worker nodes across the DC. The GM first checks its internal partitions for an available worker node. The GM's internal partitions are those partitions under each LM which have been assigned to the GM for scheduling. If an available worker node is found, the GM sends a request to the responsible LM, asking the LM to confirm that the worker node is still available and, if so, to launch the task on the available worker node. If the search is unsuccessful, the GM searches for available worker nodes in the external partitions, and sends the appropriate LM a verification and launch request. The partitions external to a GM are those that have been mapped to other GMs. This is known as a \textit{repartition} operation, in which a worker node belonging to another GM is temporarily borrowed by a GM. If there are no worker nodes available in either the internal or external partitions, the GM waits for resources to become available before scheduling the remaining tasks.

\subsection{Eventually-Consistent Global State}
Since maintaining a decentralized global state across all the GMs would require considerable synchronous communication overheads, the GMs in Megha maintain an eventually-consistent global state. Therefore, if one GM modifies its copy of the global state by marking a worker node as unavailable (on scheduling a task), the same modification may not be communicated to another GM until much later. This inconsistency in the global states stored by different GMs is addressed by the verification step in the LM. For instance, consider the following scenario where $GM\_A$ is searching for an available worker node. When it is unable to find a worker node in its internal partition it checks its global state for worker nodes in its external partitions and finds that $B1\_1$ is available. $B1\_1$, as the label indicates, belongs to $GM\_B$ in the cluster managed by $LM\_1$.  $GM\_A$ sends $LM\_1$ a request to check if the worker node is still available and can be borrowed by $GM\_A$ temporarily to run the task. $LM\_1$, which has up-to-date information on its cluster's state, finds that the worker node is available and assigns the worker node to $GM\_A$ for the duration of the task. Soon after this, $GM\_B$, which has not been informed of the unavailability of its worker node $B1\_1$, may also wish to schedule a task on the worker node. On receiving the verification request, $LM_1$, finding that the worker node is no longer available, sends $GM\_B$ an inconsistency message asking $GM\_B$ to rectify its scheduling decision. Along with this message, the LM also sends the current state information of its cluster to $GM\_B$. $GM\_B$ updates its global state with the partial state sent by the LM and revises its scheduling decision for the task.  

Since inconsistencies result in unnecessary communication and processing overheads, we have tried to reduce the inconsistencies by shuffling the worker nodes and partitions in each GM, such that the worker nodes and the partitions picked by each GM are different, thereby reducing inconsistencies. The GM also employs a round-robin algorithm when selecting the LM and partition so that tasks are distributed evenly.

\subsection{Task Completion}
When a task completes its execution, the  GM that scheduled the task and the LM to which the worker node belongs, are sent task completion messages. When the GM receives the message, it checks its job queue for pending tasks and maps a task from the queue to the now-freed worker. In the case of a task that was scheduled using repartition, on task completion, the borrower GM is only intimated that one of its tasks has completed execution, but is not allowed to schedule a pending task on the worker node. After each task completion, the scheduling GM checks if all the tasks in the corresponding job have completed execution. If yes, the job is marked as completed and removed from the job queue.

\subsubsection{Batching requests and responses}
When a job arrives at a GM, the GM picks a partition with one or more available worker nodes to schedule the job's tasks. Until the partition is saturated, that is, until it has no more available worker nodes, the GM continues to map subsequent tasks to available worker nodes in the same partition. Once the partition is saturated, the GM moves on to another partition in a round-robin fashion. Since the match operation is fairly simple and, therefore, fast, Megha batches task requests bound for the same LM and sends the multiple task-to-node mappings as a single message. For each job, the GM creates a batch of mappings of the form $<task_i,wnode_j>$, where $task_i$, is a task belonging to the job, and $wnode_j$, is the worker node chosen by the GM to launch $task_i$. The batch is limited to only those mappings with worker nodes belonging to the same LM. The LM, in turn, receives this batch of requests, iterates through them to verify it, and launches tasks for valid mappings. The LM then batches all the invalid requests (made due to inconsistencies), piggybacks an LM status update containing the current state of the LM's cluster, and sends it to the responsible GM. On receiving this batched response, the GM inserts the tasks from the invalid requests to the front of its task queue and replaces its stale version of the LM's cluster state with the updated one. 

The motivation for batching is to reduce the number of LM status updates. Updating the global state of a GM is a costly operation both in terms of processing and communication overheads. During our experiments, we found that without batching, multiple invalid requests would be made one after another due to the GM having used the same stale state to make multiple requests. If the requests and responses were not batched, the LM would receive multiple invalid requests one after another and would have to respond with its current LM state to each request. Instead, by batching requests, the LM can respond with just one update for an entire batch of invalid requests instead of an update for each invalid request made to the same LM. While the processing overhead of mapping a task to a worker node is minimal, unlimited batch sizes may lead to processing delays that may not be insignificant. Due to this, we limit the size of the batch.

\subsection{Availability}
Megha's GMs are stateless. On failure of one or more GMs, new GMs can be started and the state recovered from the heartbeats received from LMs. Additionally, the LMs also need to update the new GMs on the tasks that have been completed so far so that the GM can update its jobs store. For LM failures, traditional ways of failure recovery can be used. When a worker fails, the LM in charge can restart it; the LM must resume all tasks that were previously running on the worker using its active task list. High LM availability can be achieved using an active-standby setup, in which both the active and standby LMs get all updates from the workers and GMs, but only the active LM enforces the GMs' scheduling decisions. In our prototype, we have used the Kubernetes Master to act as Megha's LM. Kubernetes can be configured with a multi-master setup that uses redundancy to provide high availability in various failure scenarios. 

\begin{table}[htbp]
  \centering
  \caption{Workload Statistics}
  \label{tab:workload}
  \begin{tabular}{>{}p{4cm} >{\centering\arraybackslash}p{1.5cm} >{\centering\arraybackslash}p{1.5cm} >{}p{4.5cm}}
    \toprule
    Workload & \# Jobs & \# Tasks & IAT \\
    \midrule
    Yahoo trace & 24262 & 968335 & as per trace\\
    Google sub-trace & 10000 & 312558 & as per trace\\
    Synthetic trace & 2000 & 1000 & 0.025s - 0.1s (based on load) \\ 
    Down-sampled Google sub-trace & 784 & 3041 & exponentially distributed with mean 1s \\
    Down-sampled Yahoo trace & 792 & 963 & exponentially distributed with mean 1s\\
    \bottomrule
  \end{tabular}
\end{table}

\section{Evaluation Methodology}
\label{sec:methodology}

Megha was evaluated using both simulation runs and a prototype. The details of both experimental setups are as follows.

\subsection{Simulation}
Megha's event-driven simulator was built along the lines of the simulators implemented for Sparrow \cite{ousterhout2013sparrow}, Hawk \cite{delgado2015hawk}, and Eagle\cite{delgado2016job}. For consistency and better comparison, we have rewritten the simulator for Pigeon \cite{wang2019pigeon} by modifying the simulator code for Eagle. Cluster traces published by Yahoo and Google were used to study the performance of the scheduling architectures along with synthetically generated traces. The details of the jobs and tasks in the cluster traces are given in Table \ref{tab:workload}. The synthetically generated traces consisted of jobs, each with 1000 tasks of duration 1s. The inter-arrival times of the jobs were changed to achieve the required load. Load is defined as:
\begin{equation}\label{eq:load}
    Load = \frac{Resource\_Demand}{Total\_Resource}
\end{equation}
where $Resource\_Demand$ is the number of resources the workload requires per second, and $Total\_Resource$ is the number of resources available in the cluster. This equation applies to DC with a single resource or a scheduling unit. However, the equation can be used to derive the loads for different resource types as well. To study Megha, we assume a single resource unit is a scheduling unit.
DC sizes of 3,000 worker nodes and 13,000 worker nodes were configured for simulations with the Yahoo and Google cluster traces, respectively. These configurations have been borrowed from previous work \cite{delgado2016job,wang2019pigeon}. We have compared Megha with Sparrow, Eagle, and Pigeon. As in previous works, the network delay for each communication was set to a constant value of 0.5ms in all the simulation experiments. The LM heartbeat interval has been empirically determined to produce optimal results at 5s and has been set accordingly in all the simulation runs.

For runs with the synthetic traces, we vary the DC sizes between 10k-50k worker nodes. We assume the DC has been configured for peak load and therefore, we do not evaluate Megha under scenarios where the load is greater than 1.

The simulation considers communication and queuing delays, but cannot measure processing, container creation, and execution delays. Due to this limitation, we also study Megha by building and deploying its prototype on a physical cluster.

\subsection{Prototype}

We implemented prototypes of Megha and Pigeon in Python and deployed the frameworks on 3 Kubernetes clusters. The LM and the cluster coordinator components of Megha and Pigeon were implemented as web servers that communicated with the Kubernetes master of the cluster to launch tasks and receive state updates. Each Kubernetes cluster had 40 nodes with configuration 4vCPUs and 8GB RAM and one master node. We modeled each node as 4 scheduling units or worker nodes. The resultant DC consisted of 3 Kubernetes clusters with 160 worker nodes in each cluster. 

The tasks in the Google cluster sub-trace and Yahoo cluster trace used in the simulation runs were down-sampled by a factor of 100, and the inter-arrival times of the jobs were generated by modeling the job arrivals as a Poisson Process with the lambda parameter set to 1s. The workload consisted of containers that ran mathematical operations for the duration of the task execution mentioned in the trace. Since the cluster sizes are small, and the load on the DC is less than 50\%, we set the LM heartbeat interval to 10s.

\section{Results}
\label{sec:results}

The performance of Megha under different parameter settings and in comparison to other frameworks has been described in this section.

\subsection{Scalability}

\begin{figure}[htbp]
  \centering
  \begin{subfigure}[b]{0.49\textwidth}
    \centering
    \includegraphics[width=\textwidth]{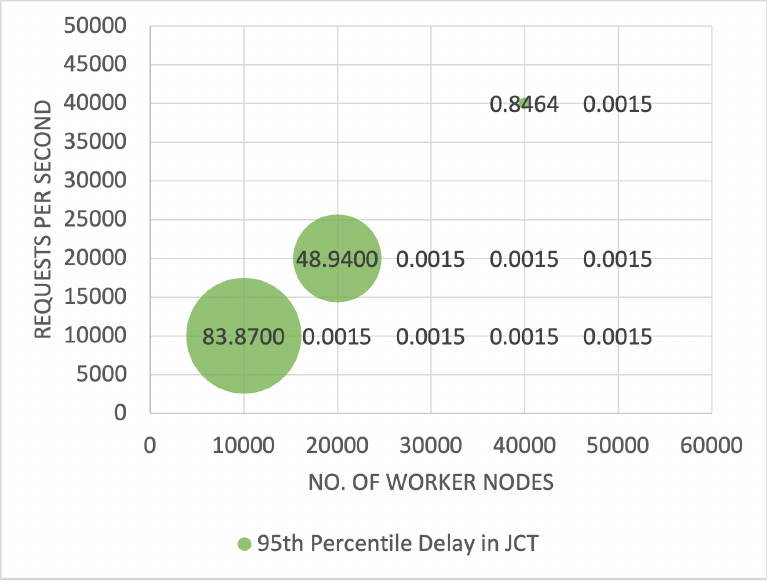}
    \caption{95th percentile of job delays under different loads}
    \label{fig:delayvsload}
  \end{subfigure}
  \hfill
  \begin{subfigure}[b]{0.49\textwidth}
    \centering
    \includegraphics[width=\textwidth]{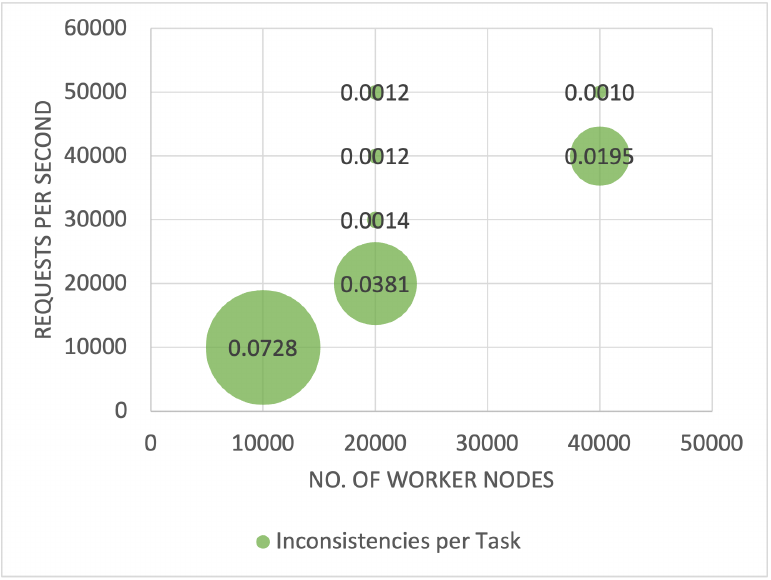}
    \caption{The ratio of inconsistencies to the total number of task requests}
    \label{fig:inconsistencies}
  \end{subfigure}
  \caption{Megha's performance under different loads}
  \label{fig:twosubfigures}
\end{figure}

We compare the number of inconsistencies and the delays experienced by the jobs scheduled by Megha under different loads. Under all loads and DC sizes, Megha delivers a median delay of 0.0015s, meaning, 50\% of the jobs do not experience queuing or additional communication delays from inconsistencies. Fig.  ~\ref{fig:delayvsload} shows the 95th percentile of the delays in JCT for different DC sizes and requests per second calculated as shown in Eq. \ref{eq:job_delay}. Fig. \ref{fig:inconsistencies} shows the number of inconsistency events per task recorded under the various loads and DC sizes. The increase in the inconsistency ratio and 95th percentile delay in jobs show that Megha's performance drops when the load on the system (calculated as the number of resources demanded divided by the number of resources available) approaches 1. This can be attributed to the increase in the GM-side queuing due to a scarcity of available worker nodes. 

\subsection{Comparison with Other Frameworks}
Megha was compared with three frameworks using simulation: Sparrow, Eagle, and Pigeon. The Yahoo cluster trace and a sub-trace of the Google cluster trace were used to evaluate the performances of the four frameworks. We also compared the performance of the prototypes implemented for Megha and Pigeon using a down-sampled version of the Google cluster trace. 

\begin{figure}[!h]
  \centering
  \begin{subfigure}[b]{0.49\textwidth}
    \centering
    \includegraphics[width=\textwidth]{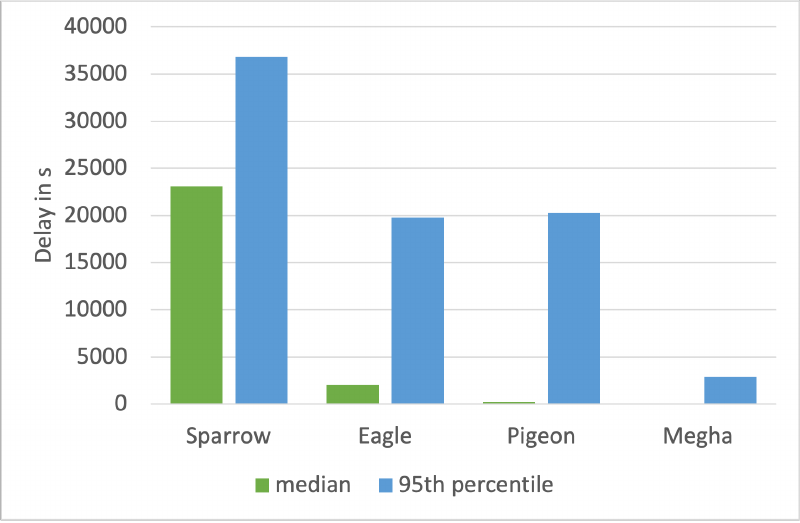}
    \caption{Delay in jobs in Yahoo trace}
    \label{fig:YH_all_jobs}
  \end{subfigure}
  \hfill
  \begin{subfigure}[b]{0.49\textwidth}
    \centering \includegraphics[width=\textwidth]{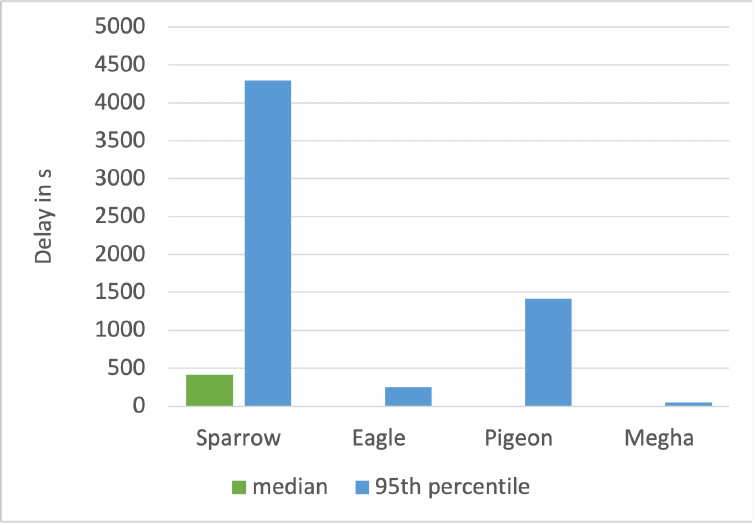}
    \caption{Delay in jobs in Google sub-trace}
    \label{fig:GOOG_all_jobs}
  \end{subfigure}
  \begin{subfigure}[b]{0.49\textwidth}
    \centering
    \includegraphics[width=\textwidth]{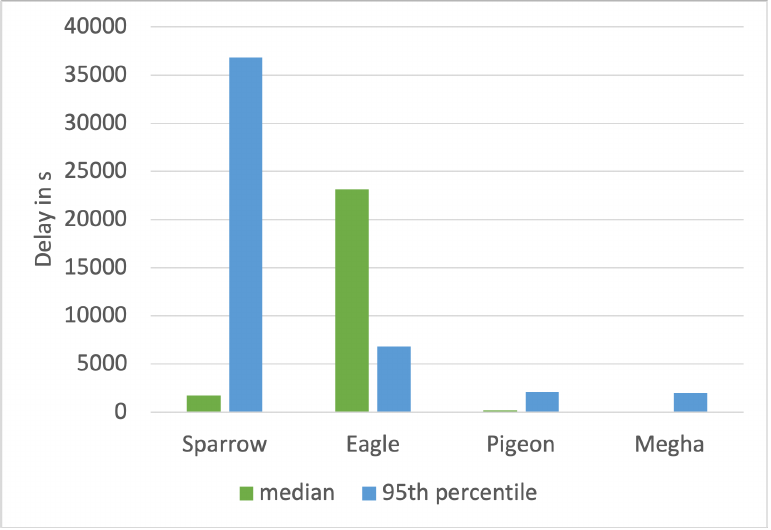}
    \caption{Delay in short jobs in Yahoo trace}
    \label{fig:YH_short_jobs}
  \end{subfigure}
  \hfill
  \begin{subfigure}[b]{0.49\textwidth}
    \centering
    \includegraphics[width=\textwidth]{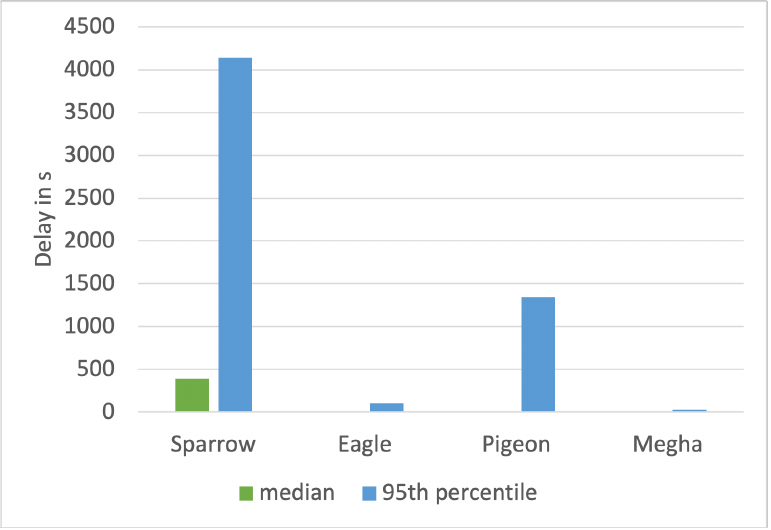}
    \caption{Delay in short jobs in Google sub-trace}
    \label{fig:GOOG_short_jobs}
  \end{subfigure}
  \captionsetup{justification=centering}
  \caption{Delays in job completion time experienced by different job types in the two workloads}
\end{figure}

The median and 95th percentile delays in JCT have been shown for the frameworks in Fig.~\ref{fig:YH_all_jobs} and \ref{fig:GOOG_all_jobs}. Megha recorded the lowest median delay under both workloads and for all job types. The delays experienced by the jobs were highest under Sparrow for all the scenarios. Even though Megha does not distinguish between short and long jobs, the delays experienced by short jobs scheduled with Megha were smaller than those experienced under Eagle and Pigeon which are priority-aware scheduling architectures. Fig.~\ref{fig:YH_short_jobs} and \ref{fig:GOOG_short_jobs} show the delays experienced by short jobs under the four frameworks. Megha reduced the average delay experienced by jobs in the Yahoo cluster trace by a factor of 12.5, 2, and  1.35 when compared to Sparrow, Eagle, and Pigeon, respectively. Megha reduced the average delay experienced by jobs in the Google cluster sub-trace by a factor of 12.89, 1.52, and  1.7 when compared to Sparrow, Eagle, and Pigeon, respectively. As explained earlier, a reduction in the overall delays in jobs directly translates to better utilization of DC resources, and therefore, higher ROI.

\subsection{Performance of Prototype}
Since the tasks scheduled by the prototype also encounter overheads from container creation, inconsistent network transfer speeds, and interference from co-located tasks, the delays reported by the prototype are higher than those reported in the simulation results. The results of the runs with Megha and Pigeon were averaged over 3 runs for each prototype and workload. Under the down-sampled Google cluster sub-trace, Megha encountered an average of 0.0015 inconsistency events per task, but under the down-sampled Yahoo cluster trace, there were no inconsistency events recorded. 

\begin{figure}[!h]
  \centering
  \begin{subfigure}[b]{0.49\textwidth}
    \centering
    \includegraphics[width=\textwidth]{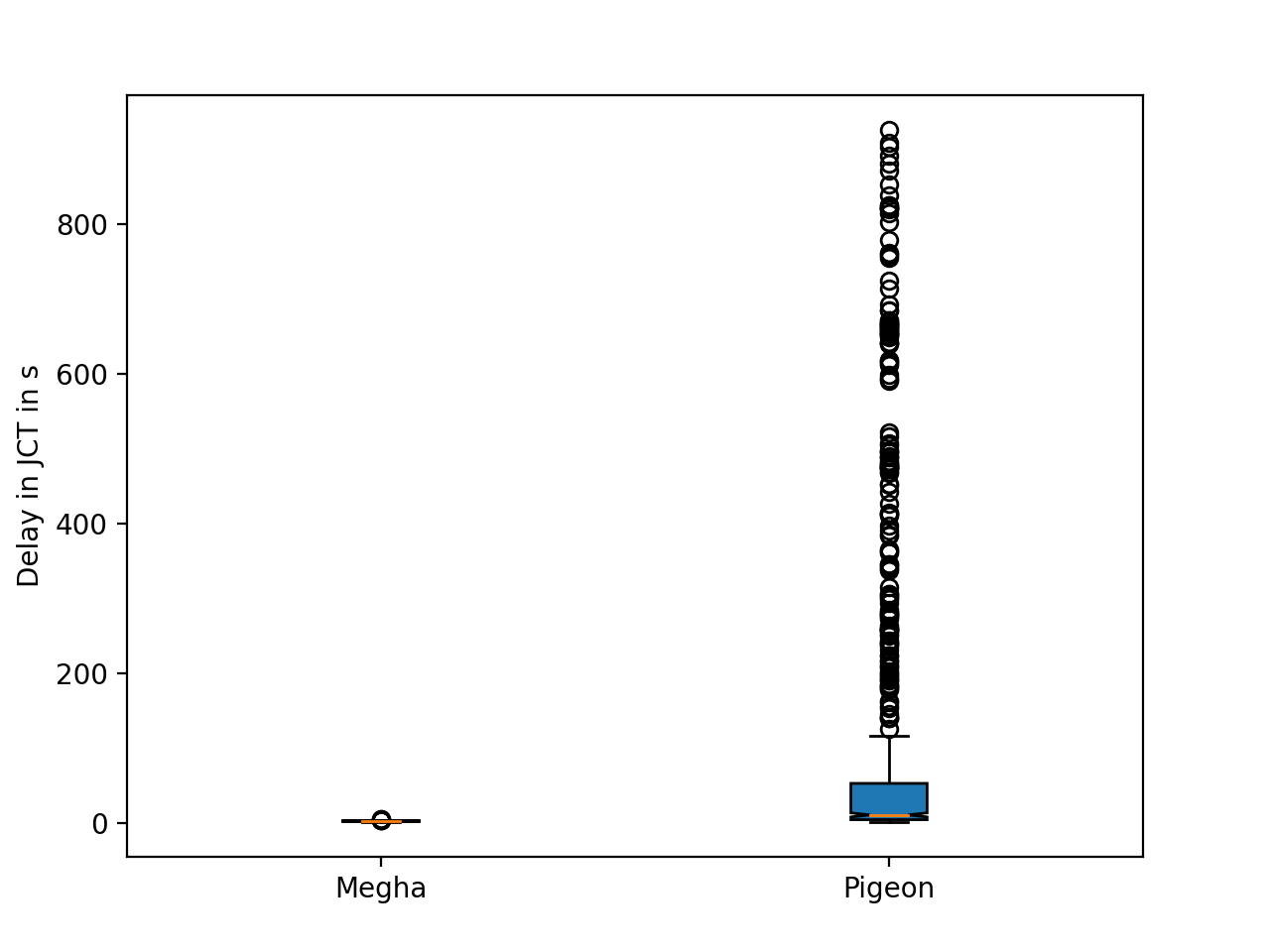}
    \caption{Delay in jobs in down-sampled Yahoo sub-trace}
    \label{fig:YH_delays}
  \end{subfigure}
  \hfill
  \begin{subfigure}[b]{0.49\textwidth}
    \centering \includegraphics[width=\textwidth]{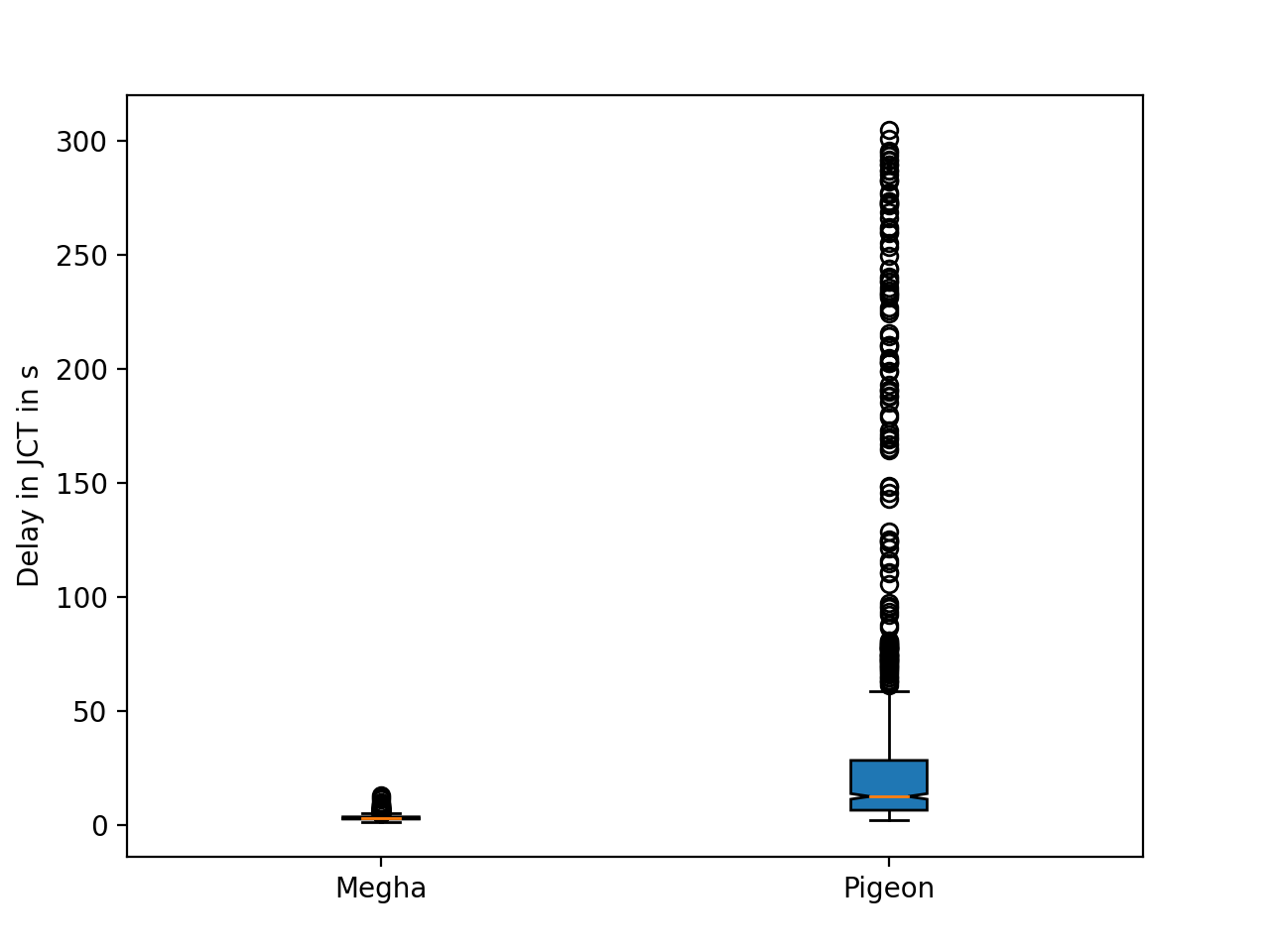}
    \caption{Delay in jobs in down-sampled Google sub-trace}
    \label{fig:GOOG_delays}
  \end{subfigure}
  \captionsetup{justification=centering}
  \caption{Delays in job completion time experienced by jobs in the two workloads}
\end{figure}

The distribution of delays in JCT experienced by jobs scheduled by the two frameworks have been shown in Fig.~\ref{fig:YH_delays} and Fig.~\ref{fig:GOOG_delays}. Jobs in both workloads have a bounded delay under Megha, whereas under Pigeon, not only is the magnitude of the delay higher, but the delay also varies considerably leading to a long tail in the delay distribution. For jobs in the Yahoo sub-trace, Megha improves on the median delay in JCT by a factor of 4 and on the 95th percentile delay by a factor of 184.5. For jobs in the Google sub-trace, Megha improves on the median delay in JCT by a factor of 4.2 and on the 95th percentile delay by a factor of 37.

\section{Related Work}
In this section, we discuss relevant research carried out in data center scheduling architectures.

PCSSampler\cite{hao2017pcssampler} is an enhancement on sample-based schedulers such as Sparrow with the help of a private cluster state, or PCS.  The PCS is created and stored by each distributed scheduler from worker node resource information sent along with probe communication. In addition to random probes, the schedulers use this cached worker node information to find available worker nodes. However, since the cached information may be stale, the scheduler may still schedule a task on to an unavailable worker node, resulting in unnecessary queuing, even though there exist available worker nodes in the cluster. Megha avoids unnecessary worker-side queuing by having the LMs verify each request sent by the GM.

Dice \cite{zhou2019improving} is a technique proposed to address the high latency experienced by short jobs when scheduled by hybrid schedulers such as Eagle\cite{delgado2016job} and Hawk\cite{delgado2015hawk}. Dice dynamically adjusts the size of the short partitions reserved to schedule latency-sensitive short jobs, thereby improving short job latency while penalizing long jobs. Additionally, it preempts long jobs when the delays in short jobs exceed a threshold. However, since short jobs are scheduled by distributed schedulers using sampling, hybrid schedulers equipped with dice still face the problem of unnecessary queuing at busy worker nodes while there are available worker nodes present in the cluster elsewhere. Megha never places a task on a busy worker node's queue. Instead, it uses schedulers equipped with a decentralized global state and verification at the LM to improve task placement decisions and reduce unnecessary queuing. However, worker reservations for short jobs are a feature we wish to evaluate with  Megha as a part of our future enhancements.

Peacock \cite{khelghatdoust2018peacock}, a probe-based scheduler, uses probe rotation to reduce worker-side queuing delays of tasks. It also uses probe reordering to ensure that jobs scheduled earlier are not delayed due to probe rotation. The framework uses an overlay network that allows worker nodes to send probes marked for rotation to their successor worker nodes. Peacock also maintains a shared state of the number of probes in the cluster and the total load in the cluster calculated using task runtime estimates and uses this to determine when rotation should be done. Though Peacock improves on traditional probe-based schedulers, it is still susceptible to queuing tasks at worker nodes that are busy even though there may be available worker nodes elsewhere in the cluster. Megha eliminates worker-side queuing by using the DC's global state and verification by the LMs to schedule tasks on available worker nodes only. It also does not use runtime estimates which may not always be available beforehand.

Fregata\cite{liu2022fregata}, is another scheduling framework that tries to optimize both resource utilization and job completion times. It has an architecture very similar to Pigeon in that it has top-level schedulers that distribute tasks in a job to multiple sub-cluster-level coordinators which map the tasks to resources. Unlike Pigeon, Fregata is dependency-aware and also tries to minimize fragmentation in nodes with its bin-packing algorithm. However, Fregata also suffers from the same problem as Pigeon wherein tasks once sent to a busy sub-cluster, cannot be migrated to available nodes elsewhere in the DC, thereby introducing unnecessary queuing overheads in jobs.

Hydra\cite{curino2019hydra} is Microsoft's proprietary federated data center scheduling architecture that separates the concerns of placement and policy to scale to a large number of worker nodes. The Hydra framework consists of multiple YARN-like sub-clusters and a global policy generator (GPG) that has access to a global state store. In each sub-cluster, a centralized Resource Manager (RM) performs monitoring, state aggregation, and container placement. The GPG determines policies and the RMs implement them in their respective sub-clusters. Due to this, the GPG is not in the critical path of task scheduling, allowing task scheduling to be performed in parallel in multiple RMs. Using a service called the AM-RM Proxy, Hydra gives the illusion of a single cluster. When a job's AM requests resources, the AM-RM Proxy routes this request to multiple RMs based on a policy, allowing a job to span multiple sub-clusters. Similarly, \textit{Routers} accept all user job requests and forward them to different RMs. When an RM receives a job request, like in YARN, it allocates resources to start a container to run the AM. We were unable to compare Megha with Hydra because neither the source code nor the workloads have been made open-source for Hydra. Megha's source code and data sets will be made open source soon. 

\section{Future Work}

Megha has no concept of worker reservations currently. We would like to study the performance gains in short jobs on introducing reservations to Megha. Additionally, we would also like to evaluate Megha's performance in scheduling tasks with placement constraints.

\section{Conclusion}
\label{sec:conclusion}
 In this paper, we present an extension of our work on Megha, a federated data center scheduler that leverages an eventually-consistent, decentralized global state to enable fast scheduling decisions. The results of our experiments showcase Megha's consistent performance across various loads and data center sizes. In order to provide a comprehensive comparison, we simulated Megha alongside other scheduling architectures. Our results reveal that, in the case of jobs within the Yahoo cluster trace, Megha achieves a reduction of the 95th percentile delays by factors of 12.5, 2, and 1.35 when compared to Sparrow, Eagle, and Pigeon, respectively. Similarly, for jobs within the Google cluster sub-trace, Megha reduces the 95th percentile delays by factors of 12.89, 1.52, and 1.7 in comparison to Sparrow, Eagle, and Pigeon, respectively.

To further validate our findings, we implemented a prototype of Megha on a 123-node Kubernetes cluster and compared its performance against a Pigeon prototype. The experimental results of our prototype deployments align with our simulation runs and demonstrate that Megha significantly improves the median delay in jobs within both the Yahoo and Google sub-traces by a factor of 4.

\section{Acknowledgements}
We would like to express our sincere gratitude to Mr. Keshava Reddy and his team from E2E Networks Limited for their invaluable support and assistance. Their generous provision of free credits has enabled us to thoroughly evaluate Megha's prototype on Kubernetes clusters.
We would also like to thank undergraduate student, Ms. Esha Arun, for her help in evaluating the performance of Megha under different data center sizes.
\bibliographystyle{elsarticle-num.bst}
\bibliography{references}




\end{document}